\documentclass[aps,groupedaddress]{revtex4}
\usepackage{graphicx}
\usepackage{amsmath}

\begin{document}

\bibliographystyle{apsrev}

\title{Reply to the comment cond-mat/0308607}

\author{H. J. Kang,$^1$ Pengcheng Dai,$^{1,2}$ 
J. W. Lynn,$^3$ M. Matsuura,$^2$ J. R. Thompson,$^{1,2}$ 
Shou-Cheng Zhang,$^4$ D. N. Argyriou,$^5$ Y. Onose,$^6$ and Y. Tokura$^{6,7,8}$}

\address{$^1$Department of Physics and Astronomy, The University of Tennessee, Knoxville, Tennessee 37996-1200}

\address{$^2$Condensed Matter Sciences Division, Oak Ridge National Laboratory, Oak Ridge, Tennessee 37831-6393 }

\address{$^3$NIST Center for Neutron Research, National Institute of Standards and Technology, Gaithersburg, Maryland 20899}

\address{$^4$Department of Physics, McCullough Building, Stanford University, Calfornia 94305-4045 }

\address{$^5$Hahn-Meitner Institute, Glienicker Str 100, Berlin D-14109, Germany}

\address{$^6$Spin Superstructure Project, ERATO, Japan Science and Technology, Tsukuba 305-8562, Japan}

\address{$^7$Correlated Electron Research Center, Tsukuba 305-8562, Japan}

\address{$^8$Department of Applied Physics, University of Tokyo, Tokyo 113-8656, Japan}

\maketitle

Mang {\it et al.} report the observation of a cubic (Nd,Ce)$_2$O$_3$ 
impurity phase grown epitaxially in annealed samples of 
electron-doped Nd$_{2-x}$Ce$_{x}$CuO$_{4}$ (NCCO) \cite{mang}.  
They claim that this impurity phase has long-range order 
parallel to the CuO$_2$ planes of NCCO but extending only 
$\sim 4a_{c}$ perpendicular to the planes, 
thus forming quasi-two-dimensional (Nd,Ce)$_2$O$_3$ lattice 
matched with the $a$-$b$ plane of NCCO.  
Although we confirmed the presence of such an impurity phase, 
(Nd,Ce)$_2$O$_3$ in our samples forms three-dimensional 
long-range structural order \cite{masato}, 
and is unrelated to the quasi-two-dimensional 
superlattice reflections~\cite{masato,kang}. 
In the paramagnetic state of (Nd,Ce)$_2$O$_3$, 
a field will induce a net moment on magnetic Nd.  
By {\it arbitrarily} scaling the impurity scattering at (0,0,2.2) 
for fields less than 7~T to our $c$-axis field-induced 
scattering of NCCO at $(1/2,1/2,0)$, 
Mang {\it et al.}~\cite{mang} argue that 
our observed magnetic scattering~\cite{kang} 
is due entirely to (Nd,Ce)$_2$O$_3$.  
This is incorrect.

There are three ways to resolve this impurity problem.  
First, if the magnetic scattering at $(1/2,1/2,0)$~\cite{kang} 
is due entirely to (Nd,Ce)$_2$O$_3$, 
one would expect that the field-induced intensity 
to be identical for ${\bf B}\parallel c$-axis and 
${\bf B}\parallel $[1,-1,0]-axis 
as required by the cubic symmetry of (Nd,Ce)$_2$O$_3$.  
Experimentally, 
we find that the field-induced effect at $(1/2,1/2,0)$ 
is much larger for ${\bf B}\parallel c$-axis~\cite{masato}, 
inconsistent with the cubic symmetry of (Nd,Ce)$_2$O$_3$ 
but consistent with the upper critical field of NCCO 
being much smaller in this geometry~\cite{masato,kang}.

Second, 
since the lattice parameter of (Nd,Ce)$_2$O$_3$ 
does not match the $c$-axis lattice parameter of NCCO~\cite{mang},
measurements at $L\neq 0$ cannot be contaminated by (Nd,Ce)$_2$O$_3$.
Our experiments indicate that the $(1/2,1/2,3)$ peak 
shows an induced antiferromagnetic (AF) component 
when the field is along the $c$-axis 
and hence superconductivity is strongly suppressed~\cite{masato},
but not when it is in the $a$-$b$ plane 
and superconductivity is only weakly affected~\cite{kang}. 
This is {\it direct} and {\it unambiguous} proof of the connection between 
field-induced AF order and suppression of superconductivity in NCCO.
We also note that the qualitatively different behavior 
observed for ${\bf B}\perp c$ vs. ${\bf B}\parallel c$-axis directly violates 
the cubic symmetry of (Nd,Ce)$_2$O$_3$.

Finally, 
an independent report by Fujita {\it et al.}~\cite{fujita} 
confirmed the major findings of Refs. 2 and 3 in studies of 
annealed superconducting Pr$_{0.89}$LaCe$_{0.11}$CuO$_{4}$ (PLCCO), 
a similar electron-doped material 
where the cubic impurity phase (Pr,La,Ce)$_2$O$_3$ 
has a nonmagnetic ground state and no field dependence 
below 7~T~\cite{kang2}.  
For fields up to 5~T, 
Fujita {\it et al.}~\cite{fujita} found enhanced AF order 
at $(1/2,3/2,0)$ with increasing field in PLCCO.  
Above 5~T, this AF order decreases with increasing field, 
consistent with the field dependence of $(1/2,1/2,0)$ in NCCO~\cite{kang}.
The overwhelming agreement on two different electron-doped systems 
found by two independent experiments~\cite{masato,kang,fujita} 
strongly confirms the quantum phase transition 
from the superconducting to an AF state 
in electron-doped high-$T_{c}$ superconductors~\cite{kang}.

\end{document}